\documentclass[sigconf]{acmart}
\AtBeginDocument{%
  \providecommand\BibTeX{{%
    Bib\TeX}}}

\setcopyright{acmlicensed}
\copyrightyear{2024}
\acmYear{2024}
\acmDOI{XXXXXXX.XXXXXXX}

\acmConference[ICCAD '24]{ACM/IEEE International Conference on Computer-Aided Design}{October 27--31,
  2024}{Newark, NJ, USA}
\acmBooktitle{ICCAD '24: ACM/IEEE International Conference on Computer-Aided Design,
  October 27--31, 2024, Newark, NJ}
\acmISBN{978-1-4503-XXXX-X/18/06}

\usepackage{amsmath}
\usepackage{mathtools}
\usepackage{amsfonts}
\usepackage{caption}
\usepackage{algorithmic}
\usepackage[ruled,vlined]{algorithm2e}
\usepackage{graphicx}
\usepackage{textcomp}
\usepackage{xcolor}
\usepackage{verbatim}
\usepackage{subcaption}
\usepackage{adjustbox}
\usepackage{booktabs}
\usepackage{multirow}
\usepackage{threeparttable, tablefootnote}
\usepackage{varwidth,xcolor}

\usepackage{tabularray}
\usepackage{ragged2e}
\usepackage{mathtools}
\usepackage{array}

\newcommand{\ignore}[1]{}
\newcommand{\fixme}[1]{\textcolor{red}{#1}}

\newcommand{\norm}[1]{\left\lVert#1\right\rVert}

\copyrightyear{2024}
\acmYear{2024}
\setcopyright{rightsretained}
\acmConference[ICCAD '24]{IEEE/ACM International Conference on Computer-Aided Design}{October 27--31, 2024}{New York, NY, USA}
\acmBooktitle{IEEE/ACM International Conference on Computer-Aided Design (ICCAD '24), October 27--31, 2024, New York, NY, USA}
\acmDOI{10.1145/3676536.3676832}
\acmISBN{979-8-4007-1077-3/24/10}

\begin{document}
\newcolumntype{C}[1]{>{\centering\arraybackslash}p{#1}}
\title[Hyena: Optimizing Homomorphically Encrypted Convolution for Private CNN Inference]{Hyena: Optimizing Homomorphically Encrypted Convolution \\for Private CNN Inference}


\author{Hyeri Roh and Woo-Seok Choi}
\affiliation{%
  \institution{Dept. of ECE, ISRC, Seoul National University}
  \streetaddress{1 Gwanak-ro, Gwanak-gu}
  \city{Seoul}
  \country{South Korea}}
  \email{{hrroh,wooseokchoi}@snu.ac.kr}


\begin{abstract}
Private inference using homomorphic encryption has gained a great attention to leverage powerful predictive models, e.g., deep convolutional neural networks (CNNs), in the area where data privacy is crucial, such as in healthcare or medical services.
Processing convolution layers, however, occupies a huge portion (more than 85\,\%) of the total latency for private CNN inference.
To solve this issue, this paper presents Hyena utilizing a novel homomorphic convolution algorithm that provides speedup, communication cost, and storage saving.
We first note that padded convolution provides the advantage of model storage saving, but it does not support output channel packing, thereby increasing the amount of computation and communication.
We address this limitation by proposing a novel plaintext multiplication algorithm using the Walsh-Hadamard matrix.
Furthermore, we propose the optimization techniques to significantly reduce the latency of the proposed convolution by selecting the optimal encryption parameters and applying lazy reduction.
Overall, Hyena achieves 1.6-3.8$\times$ speedup and reduces the weight storage by 2000-8000$\times$ compared to the conventional convolution.
\ignore{Specifically, for networks like VGG-16 and ResNet-20 on ImageNet, and ResNet-18 on Tiny ImageNet, it boasts a latency 1.21-2.55$\times$ faster and a memory usage reduced by 2.1-7.9$\times$ relative to conventional networks.}
For deep CNNs like VGG-16, ResNet-20, and MobileNetV1 on ImageNet, Hyena reduces the end-to-end latency by {1.3-2.5}$\times$, the memory usage by 2.1-7.9$\times$ and communication cost by {1.4-1.5}$\times$ compared to conventional method.

\ignore{
The growing concern about the protection of user data has led to the increasing significance of private convolutional neural networks (CNNs) for inferences in the cloud.
Homomorphic encryption (HE) is a potential solution for this issue, as it enables cloud services to perform inferences directly on clients' encrypted data in big-data deep learning.
However, this approach poses significant computational challenges and remains impractical in current systems.
To address this,} 
\ignore{We present a novel convolution methodology with parameter and algorithmic optimizations for server-side HE CNN inferences.}
\ignore{This methodology involves performance modeling to determine the optimal parameters for constructing an optimized CNN that balances both latency and memory usage. 
The average error of performance modeling is 4\% in the baseline scheme, while the average error of the optimized schemes is 1.3\% for VGG16 and ResNet20/18.}
\ignore{It boosts the 1.64-5.43$\times$ faster offline latency with 1.33-1.55$\times$ reduced offline client-to-server communication costs compared to prior work.
During the online phase, they are 1.21-2.55$\times$ faster in latency and 1.7-1.94$\times$ reduced communication cost.
Furthermore, memory usage is reduced by 2.1-7.9$\times$ compared to the baseline.}
\ignore{Together with optimization, it reduces latency in linear layers and saves memory consumption in the overall network.
they are 1.21-2.55$\times$ faster in latency.
Furthermore, memory usage is reduced by 2.1-7.9$\times$ compared to the baseline.}

\end{abstract}
\maketitle
\section{introduction}
\label{sec:intro}

In recent years, many machine learning services including classification, prediction, and recommendation, run on clouds to leverage powerful hardware and software resource rather than running them on local devices.
However, this cloud-based inference comes with significant data privacy concerns, which is especially crucial for industries where data sensitivity is high but the potential benefits of machine learning are significant such as in healthcare and financial services.
To mitigate such privacy issues, there has been active research on developing cryptographic and algorithmic techniques that allow the service users or clients to obtain inference results from the models without revealing the input data to the clouds hosting the models.

\ignore{
Machine Learning as a Service (MLaaS) has emerged as a popular solution for providing clients with access to cloud-based inferences without the need for them to perform large computations on their own. 
However, this has also brought to light concerns regarding the security of sensitive data when it is transmitted to servers for processing.
To address this issue, privacy-preserving machine learning (PPML) inference is gaining traction as a means of ensuring the privacy and security of both the client's data and the server's trained models~\cite{gentry2009fully}.
}

Several techniques such as homomorphic encryption (HE) or secure multi-party computation (MPC) can be used for such private inference (PI), each with their distinct pros and cons in terms of computation/communication overhead or inference accuracy.
For instance, PI protocols using HE only process the entire network on the server side while replacing the nonlinear layers like ReLU with low-degree polynomials to reduce computational overhead~\cite{gilad2016cryptonets,brutzkus2019low}. 
However, altering nonlinear layers to low-degree polynomials sacrifices inference accuracy~\cite{garimella2021sisyphus}.

\begin{figure}[t]
\centering
\includegraphics[width=0.9\columnwidth]{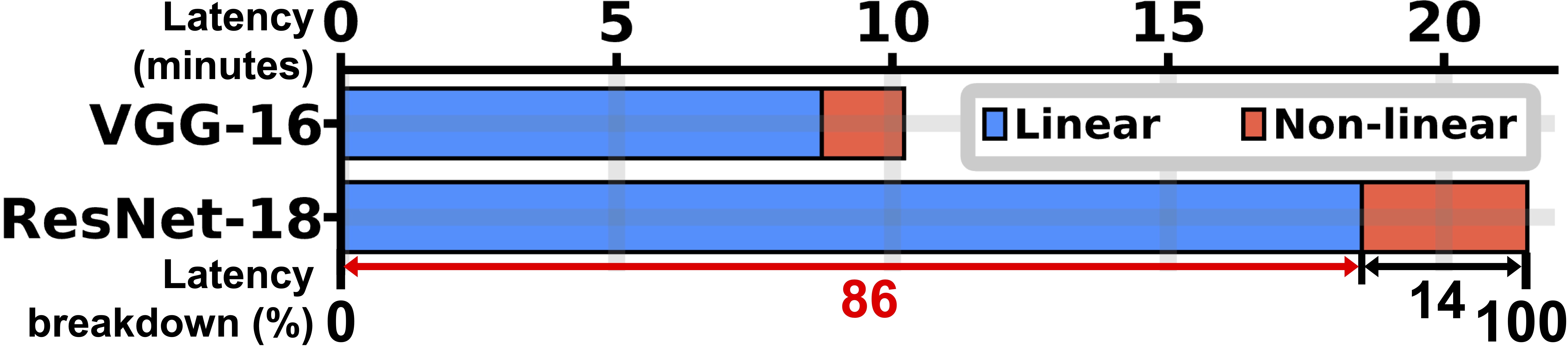}
\vspace{-0em}
\caption{\ignore{All-ReLU Delphi protocol: (a)} Latency breakdown between linear vs. nonlinear layers for PI with VGG-16 and ResNet-18 on TinyImagenet~\cite{garimella2023characterizing}. Processing convolution layers occupies more than 85\,\% of the total end-to-end latency. \ignore{and online (on.) vs offline (off.) phases. (b) The amount of pre-processing data for VGG-16 on ImageNet: ReLU for the non-linear layer and model weights for the linear layers.}}
\vspace{-1em}
\label{breakdown}
\end{figure}

Hybrid PI protocols have been proposed to compute linear layers using HE and non-linear layers using MPC~\cite{juvekar2018gazelle,reagen2021cheetah,mishra2020delphi}. 
\ignore{These protocols assume that the server and client are semi-honest, and they often use Yao's garbled circuits~\cite{yao1986generate} and secret sharing to process the ReLU layer.
This paper focuses on using HE only in convolution layers, assuming the use of hybrid methods, which can be faster than HE-only without any loss of accuracy, to provide privacy-preserving deep learning inference.
They often use Yao's garbled circuits~\cite{yao1986generate} and secret sharing to process the ReLU layer.}
They do not alter the pretrained model architecture, thereby preventing any accuracy loss.
However, state-of-the-art protocols like Delphi~\cite{mishra2020delphi} suffer from two main drawbacks:
1) as indicated in Fig.~\ref{breakdown}, latency to process linear layers dominates that of nonlinear layers for deep convolutional neural networks (CNNs),
and 2) it requires offline phase during which computation and communication between client and cloud need to be performed for every inference.
According to \cite{garimella2023characterizing}, if the offline phase computation time is prolonged, it subsequently impacts the online phase latency. 
\ignore{Therefore, in terms of latency, both offline and online phases are crucial.}
Therefore, it is critical to develop a low-latency HE convolution algorithm that does not require any offline phase to improve PI performance.

There exist HE convolution algorithms proposed in \cite{juvekar2018gazelle,reagen2021cheetah,choi2022impala,ran2023spencnn} that do not need offline phase, but they also suffer from drawbacks.
Since \cite{choi2022impala,ran2023spencnn} exploit model sparsity to reduce computation, they either suffer from accuracy degradation or cannot be applied to general non-sparse models.
For \cite{juvekar2018gazelle,reagen2021cheetah}, model weights for convolutional layers must be stored in polynomial form (plaintext), which requires model storage capacity excessively.
\ignore{, while non-linear layers require ReLU operations using garbled circuits~\cite{juvekar2018gazelle, reagen2020cheetah, mishra2020delphi, garimella2023characterizing}.
We adopt the all-ReLU Delphi~\cite{han2015deep} protocol as our baseline.}
For instance, the VGG-16 model on ImageNet consumes more than 216\,GB of storage.
This huge memory\ignore{storage} capacity requirement degrades the convolution latency especially when co-locating many models (i.e., processing many HE convolutions in parallel) on a single machine for higher throughput in data centers.

In view of these drawbacks, we propose a low-latency low-storage HE convolution without offline phase that can be applied to general convolution operators, e.g., standard, grouped, depth-wise convolution, etc., without accuracy degradation.
Specifically, this paper makes the following contributions:
\begin{itemize}
\item We introduce Hyena, a novel HE convolution algorithm using the Walsh-Hadamard matrix, which enables output channel packing to minimize the output ciphertexts as well as model weight storage saving. Hyena reduces the weight storage by 2000-8000$\times$ by eliminating the need to store weights in plaintext.
\item Hyena incorporates latency optimization techniques, leveraging optimal encryption parameter selection for minimal noise growth and a lazy reduction strategy to minimize costly HE operations.  Compared to the conventional method, Hyena achieves 1.6-3.8$\times$ speedup.
\ignore{\item We present a mean absolute error of 5.45\,\% for the proposed convolution modeling and 3.76\,\% for the conventional convolution modeling. This modeling yields an optimal set of parameters harmonizing latency and memory consumption, facilitating the design of CNNs across a broad spectrum of parameter sets.}
\ignore{\item We implemented VGG-16 and ResNet-20 on ImageNet, and ResNet-18 on TinyImageNet. This resulted in memory savings between 2.1-7.9$\times$ when considering both key and model sizes, and achieved a latency reduction of  1.2-2.6$\times$ for linear layers.}
\ignore{\item \fixme{We study performance characteristics of running homomorphic convolution operators in practical datacenter environments and show that state-of-the-art techniques such as \cite{} are susceptible to latency degradation when co-locating multiple operators on a single machine.}}
\item We apply Hyena to VGG-16, ResNet-20, and MobileNetV1 on ImageNet and evaluate their performance. Hyena achieves 2.1-7.9$\times$ total memory saving, {1.3-2.5}$\times$ speedup, and {1.4-1.5}$\times$ communication cost saving.
\item We study the performance sensitivity of running multiple HE convolutions on a single machine and demonstrate their impact on the cache miss rate. We show that Hyena shows the least latency variation compared to other convolution techniques such as Gazelle~\cite{juvekar2018gazelle} and Cheetah~\cite{huang2022cheetah}.

\ignore{\item \fixme{End-to-end implementation. Communication cost.}}
\end{itemize}

\section{Background}
\label{sec:background}

\subsection{Threat Model}

We focus on a privacy-preserving CNN inference system involving two parties, i.e., cloud and client.
Specifically, cloud performs inference on client's sensitive data while ensuring that individual private data is not revealed to cloud,
while no information about the cloud's proprietary CNN model parameters, e.g., weights, leaks to client.
However, the model architecture or hyperparameters such as the number of layers, the type of layers, or input/output dimensions are known to both parties.

As in prior art~\cite{choi2022impala,gilad2016cryptonets,huang2022cheetah,juvekar2018gazelle,mishra2020delphi,ran2023spencnn,rathee2020cryptflow2,roh2024flash}, we assume that both parties are ``semi-honest.''
In other words, both cloud and client follow the protocol honestly, 
but they may try to learn additional information about the data or model beyond what is explicitly allowed by the protocol.

\subsection{Hybrid PI Protocol}
\ignore{We employ the MPC framework, like Gazelle~\cite{juvekar2018gazelle} and Cheetah~\cite{reagen2020cheetah}, to structure the entire network.
The server executes linear layers while both client and server collaborate on non-linear layers.
During non-linear computation, they communicate using a ``garbled circuit" that supports all boolean operations, enabling lossless execution of the ReLU function at the cost of communication. 
The key insight here is that the entire process retains accuracy as long as the linear phase remains lossless.
Given that the proposed method preserves all filter values, conventional and proposed implementations avoid loss of accuracy.}

In this work, we assume that the hybrid PI protocol such as \cite{juvekar2018gazelle,reagen2021cheetah,choi2022impala,rathee2020cryptflow2,huang2022cheetah,roh2024flash}, where linear and nonlinear layers are processed with HE and MPC (e.g., arithmetic secret sharing, garbled circuit, etc.), respectively, is employed. 
The hybrid protocol provides the advantage that, once the neural network with fixed-point quantization is trained, 
we can convert and use it for PI without any accuracy degradation,
obviating the need for model retraining.
It also does not require computationally costly bootstrapping operation.
However, in state-of-the-art PI protocols, the computational overheads of HE convolution as pointed out in Fig.~\ref{breakdown} remains a huge bottleneck, especially for deep CNN models.
Therefore, this paper focuses on addressing the issues associated with HE convolution.

\subsection{Homomorphic Operations in BFV}

The BFV (Brakerski/Fan-Vercauteren) encryption scheme~\cite{fan2012somewhat} is adopted in Hyena, where a data vector is encoded into a plaintext $m$ using batch encoding for single-instruction-multiple-data (SIMD) processing~\cite{smart2014fully}, and the plaintext $m$ is represented as a polynomial of degree $n-1$ with coefficients of integers modulo $p$ (plaintext modulus). 
This plaintext is then encrypted using a secret key with random noise to produce a ciphertext $\llbracket m\rrbracket$, which comprises two polynomials ($c[0], c[1]$), each with degree $n-1$ and coefficients of integers modulo $q$ (ciphertext modulus).
The noise embedded in a ciphertext accumulates after each operation applied on the ciphertext and when it exceeds a certain noise margin (decided by the ratio $q/p$), decryption of the resulting ciphertext fails, yielding a wrong result.
Increasing $q$ expands this noise margin but incurs more computational overhead. 
Similarly, reducing $p$ also improves the noise margin at the expense of data precision.
For the following operations, consider two plaintexts $m_0$ and $m_1$, 
producing the ciphertexts $\llbracket m_0\rrbracket$=$(c[0], c[1])$ and $\llbracket m_1\rrbracket$=$(c'[0], c'[1])$, respectively.

\subsubsection{Addition}

By adding each ciphertext polynomial coefficient-wise, two ciphertexts can be added.\ignore{If we pack the multiple data into a single ciphertext, this process is analogous to SIMD addition~\cite{smart2014fully}.}
After this HE addition, noise grows additively.
\[
\textbf{HAdd}(\llbracket m_0\rrbracket,\llbracket m_1\rrbracket){=}\llbracket m_0\text{+} m_1\rrbracket{=}(c[0]+c'[0], c[1]+c'[1])
\]

\subsubsection{Multiplication}
\ignore{
There are two instances of homomorphic multiplication: ciphertext-ciphertext and ciphertext-plaintext. 
Since the cloud has its model parameters in plaintext, only the ciphertext-plaintext multiplication is used to construct the PPML convolution. 
If we pack the data into a single ciphertext, this procedure is comparable to SIMD multiplication, and the noise increases multiplicatively when performing it. 

Note that the multiplicative factor is dependent on the coefficients’ magnitude and sparsity of the multiplied plaintext polynomial, but even when slightly different values or small values are packed in plaintext, the coefficient magnitude can be as large as $p/2$, and the polynomial is not sparse at all. Therefore, generally, plaintext multiplication with packing incurs the multiplicative factor of $\sqrt{n}p/2$ [7], and storing all the plaintexts corresponding to the model parameters requires a large storage size. 
}

In PI, since cloud possesses model weights either in plaintext or scalar format, 
only ciphertext-plaintext multiplication ({\bf PMult}) and ciphertext-scalar multiplication ({\bf CMult}) are utilized. 
\begin{gather}
\textbf{PMult}(\llbracket m_0\rrbracket,m_1)=\llbracket m_0 m_1\rrbracket=(c[0]\times m_1, c[1]\times m_1) \nonumber \\
\textbf{CMult}(\llbracket m_0\rrbracket,a)=\llbracket am_0\rrbracket=(c[0]*a, c[1]*a) \nonumber
\end{gather}
$\times$ denotes coefficient-wise multiplication, whereas $*$ represents the multiplication of all coefficients by the constant $a$ simultaneously.
{\bf PMult} is used when cloud wants to multiply a weight vector element-wise with the encrypted data vector ($m_1$ SIMD-encodes the weight vector to be multiplied), whereas {\bf PMult} is used when the constant $a$ needs to be multiplied with all the elements in the encrypted data vector.
Noise increases multiplicatively after HE multiplication;
$\textbf{PMult}$ boosts noise by a factor of $np/2$ in the worst case,
while $\textbf{CMult}$ does by only a factor of the scalar $a$~\cite{choi2022impala}.

\subsubsection{Rotation}
\ignore{While addition and multiplication allows operations slot-wisely, slot rotation needs to come first to align the ciphertext with the proper slots in order to compute the data between different slots. 
The rotation operation~\cite{brakerski2014leveled,wu2012using} involves a key-switching operation that is a computationally demanding process that additively increases noise. 
As the noise from a single rotation may exceed the allowed noise budget, we have to decompose the ciphertext before the permutation operation. 
The permutation decomposition base is carefully chosen to be a manageable size because it determines both the amount of pieced ciphertext and the noise growth for the rotation of ciphertext. 
To achieve the best performance, the decomposition base should be selected as large as possible to produce little ciphertext and computation. 
In contrast, when we reduce the decomposition base, the amount of decomposed ciphertext grows, increasing the amount of computation and operation execution time. 
However, if we use the small base, the noise that the rotation causes is small, so we can save more on the noise budget and use other operations.}

When arithmetic operations between different slots in an encrypted vector are desired, rotation is necessary for data realignment.
\[
\textbf{HRot}(\llbracket m_0\rrbracket,step)=\llbracket\langle m_0\rangle_{step}\rrbracket
\]
$\langle m_0\rangle_{step}$ denotes the left-cyclic shift of the slots in $m_0$ by $step$.
The rotation operation is computationally costly due to the sub-routine {\it key-switching}, which introduces additional noise.
Managing this noise is pivotal, and the ``decomposition'' process to break down polynomials into smaller coefficient segments is central to it.
As the decomposition base is reduced, more decomposed polynomials are created, which curbs noise growth at the expense of increased computation and storage overhead for key-switching public keys. 
Conversely, a larger base leads to reduced computations and key sizes at the cost of increased noise.
\ignore{A larger base reduces computation time but increases noise, while a smaller one does the opposite.}
Thus, selecting the optimal base that minimizes the computational overhead while limiting the noise within the noise margin for correct computation is crucial.

When rotating the same ciphertext multiple times, the ``hoisting'' technique~\cite{juvekar2018gazelle} provides a speed-up.
This technique splits the rotation into two distinct steps:
1) \ignore{\textbf{permDecomp}: }decomposing the ciphertext based on the chosen base, and
2) \ignore{\textbf{permAuto}: }executing the actual rotation using the decomposed pieces.
In hosting, the first step, which is more costly, is executed just once for a given ciphertext and its result is reused for the subsequent rotations, and only the second step is performed independently multiple times.
It significantly enhances computational efficiency.

\ignore{
\begin{figure}[t]
\centering
\includegraphics[width=\columnwidth]{Fig/convolution5.png}
\caption{
Single-input single-output convolution}
\label{convolution}
\end{figure}

\begin{figure*}[t]
\centering
\includegraphics[width=\columnwidth]{Fig/gazelle6.png}
\caption{Convolution of Gazelle. The first subscript denotes the output channel, and the second does the input channel: (a) Multiple input and output channels with packing where f denotes filter, (b) Example of getting partial sum when filter $= 3*3$ with hoisting technique}
\label{gazelle}
\end{figure*}

\begin{figure*}[t]
\centering
\includegraphics[width=\columnwidth]{Fig/padded4.png}
\caption{Padded convolution. The superscript with $f$ denotes the order of the filter scalar. The first subscript denotes the output channel, and the second does the input channel: (a) Multiple input and output channels with packing, (b) Example of getting partial sum when filter $= 3*3$ with hoisting technique}
\label{padded}
\end{figure*}
}

\begin{figure*}[t]
\centering
\includegraphics[width=0.85\textwidth]{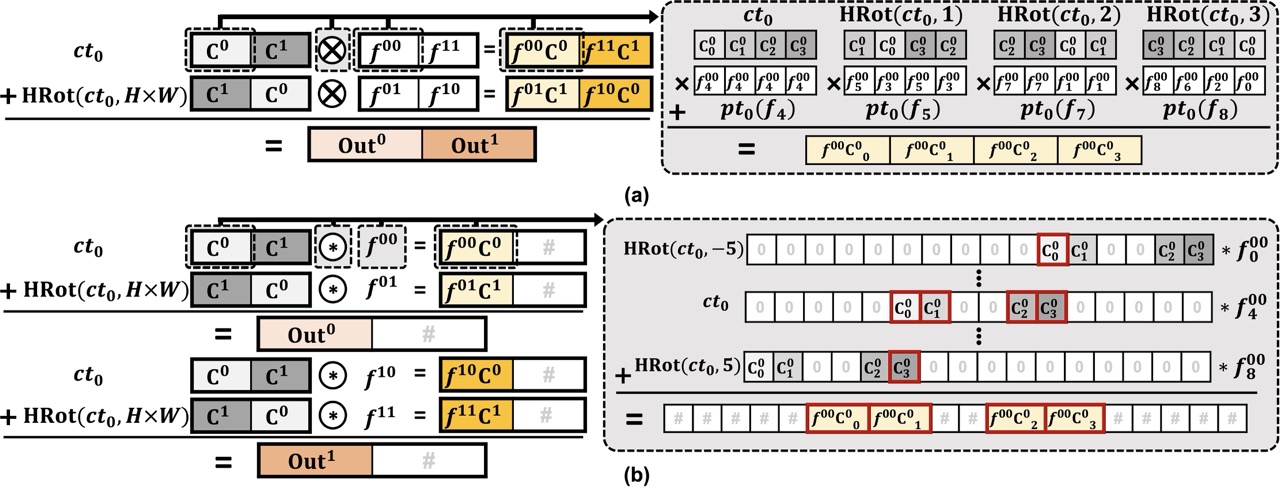}
\vspace{0em}
\caption{HE convolutions with multiple input and output channels: (a) conventional (packed) and (b) padded convolution. For $f$, the subscript denotes the weight scalar order, while the superscripts specify the output and input channels, respectively.}
\vspace{0em}
\label{conventional}
\end{figure*}

\subsection{Prior Art on Homomorphic Convolution}\label{subsec:conv}

Convolution operates on a three-dimensional tensor with dimensions: channel ($C$), height ($H$), and width ($W$) of the feature map.
To maximize slot utilization (i.e., to minimize the ciphertext size and communication cost), the $C\times H\times W$ input vector gets divided into $\lceil \frac{C\times H\times W}{n} \rceil$ ciphertexts. 
Here, $\lceil \rceil$ indicates the ceiling function, and $n$ represents the number of slots available in a plaintext. 
Depending on the size of $n$ and $H\times W$, a single ciphertext can pack multiple channels, or when $n<H\times W$, multiple ciphertexts are needed for a single channel.

For conventional convolution using SIMD encoding such as Gazelle~\cite{juvekar2018gazelle} and its variants~\cite{choi2022impala,rathee2020cryptflow2}, to maximize the slot utilization, input feature maps are encrypted without padding zeros at the boundary, and the kernel elements are packed with zeros in plaintexts, aligning with the appropriate input feature map.
Take, for example, the scenario illustrated in Fig.~\ref{conventional}(a),
where a $2\times 2\times 2$ input is convolved with four $3\times 3$ filters with elements labeled $f_0, ..., f_8$ to produce a $2\times 2\times 2$ output.
The first ($\mathbf{C^0}$) and second ({$\mathbf{C^1}$}) input channels are packed into a single ciphertext labeled ${ct_0}$.
Kernels, denoted as $\boldsymbol{f^{out\ ch., in\ ch.}}$, are packed into the plaintext $pt_0$ interleaved, alternating between input and output channels.
The convolution operation then involves computing the sum of
$\textbf{PMult}(\textbf{HRot}(ct_0, i), pt_0(f_{i+4}))$ for $i\in [-4,4]$, where $pt_0(\cdot)$ represents a rotated version of $pt_0$ aligning the filter with the input.
The hoisting technique can be applied here, as the same $ct_0$ undergoes multiple rotations. 
This process is repeated to get partial sums, which are then $\textbf{HAdd}$ed to produce the output ciphertext packing $\mathbf{Out^0}$ and $\mathbf{Out^1}$.

On the other hand, the padded convolution~\cite{juvekar2018gazelle} pads zeros to the input channels $\mathbf{C^0}$ and $\mathbf{C^1}$ rather than putting zeros in the filter plaintext slots. 
Although this slightly degrades slot utilization, eliminating the need to pad zeros in the kernel slots enables implementing single-channel convolution using {\bf CMult} instead of {\bf PMult}.
Furthermore, since the filters do not need to be prepared as plaintexts for PI, the padded convolution allows for the storage of filters as scalars rather than plaintexts. 
As a result, each kernel element can be stored as an integer, not as a plaintext, saving the storage by a factor of up to $n$.
By using $\textbf{CMult}$ $3\times 3$ times, as opposed to using $\textbf{PMult}$, it derives the partial sum as depicted in Fig.~\ref{conventional}(b).
While the subsequent process to generate the output is similar to conventional convolution, a significant distinction arises: it is impossible to pack the output channels, increasing the communication cost in the PI protocol, 
and the required number of HE multiplication increases because each output channel should be processed separately. 
However, we should note that, since $\textbf{CMult}$ increases noise much less than $\textbf{PMult}$, 
the padded convolution allows choosing larger decomposition base for $\textbf{HRot}$, 
thereby reducing the computational overhead for rotation.
Hence, the total amount of computation for the padded convolution can remain similar to that of the conventional packed convolution. 
In summary, the padded convolution results in 1) much less model weight storage requirement, 2) similar latency, and 3) higher communication cost than the conventional convolution. 
In view of this, this paper propose Hyena, a novel HE convolution that has both low model weight storage and low communication cost, as well as having less latency through several optimization techniques.  

There exist other HE convolution techniques such as Cheetah~\cite{huang2022cheetah,xu2023falcon} that exploit non-SIMD-based encoding schemes to avoid expensive rotation ({\bf HRot}), 
which can provide a large speed-up. 
However, such non-SIMD-based HE convolutions also have similar drawbacks: 
1) they require large model storage capacity since the model weights need to be packed in the plaintext format, and
2) they result in slot under-utilization during computation and do not support output channel packing, which incurs high communication cost due to the large number of output ciphertexts.
The optimization technique described in \cite{xu2023falcon} to reduce the communication cost for the non-SIMD-based HE convolution is only applicable to depthwise convolution and does not reduce communication for standard convolution. 
As will be described in Section~\ref{sec:results}, Hyena provides lower convolution latency than non-SIMD-based techniques thanks to its lower communication cost.
\section{Proposed Convolution}
\label{sec:proposed}

We propose Hyena that leverages the low-storage benefit of padded convolution, while addressing its limitation associated with output channel packing.
Additionally, techniques based on optimal encryption parameter selection and lazy reduction to reduce latency of Hyena are introduced.

\subsection{Enabling Output Channel Packing}

\ignore{
Assume that the polynomial degree $n$ is 2048 and the input size and the kernel size are $32\times32\times2$ and $3\times3\times2$, respectively. 
In the case of Gazelle, two channels are packed in a single ciphertext, and convolutions for both channels are performed simultaneously.
Storing the plaintext polynomials corresponding to the $3\times3\times2$ kernels requires 288\,KB.
Note that the multiple output channels can also be packed for the padded convolution by packing multiple kernel parameters in a kernel plaintext; however, this needs the same amount of memory storage space (288\,KB) as Gazelle.
For a single output channel, the main difference from the multiple output channels is that Gazelle places the corresponding kernel values in plaintext slots, while the padded convolution fills all of the plaintext slots with the identical value.
When all values in the plaintext slots are the same, the corresponding plaintext polynomial is reduced to a single scalar, i.e. zero-th order polynomial, greatly saving the required storage.
So, for the case of a single output channel, the padded convolution requires only 144\,B to store the kernel parameters, but the amount of computation increases because each output channel should be processed separately. 
To get around this issue, we need to create a novel scalar that makes use of padded convolution and can even provide multi-packed outputs.
}
\begin{figure}[t]
\centering
\includegraphics[width=0.85\columnwidth]{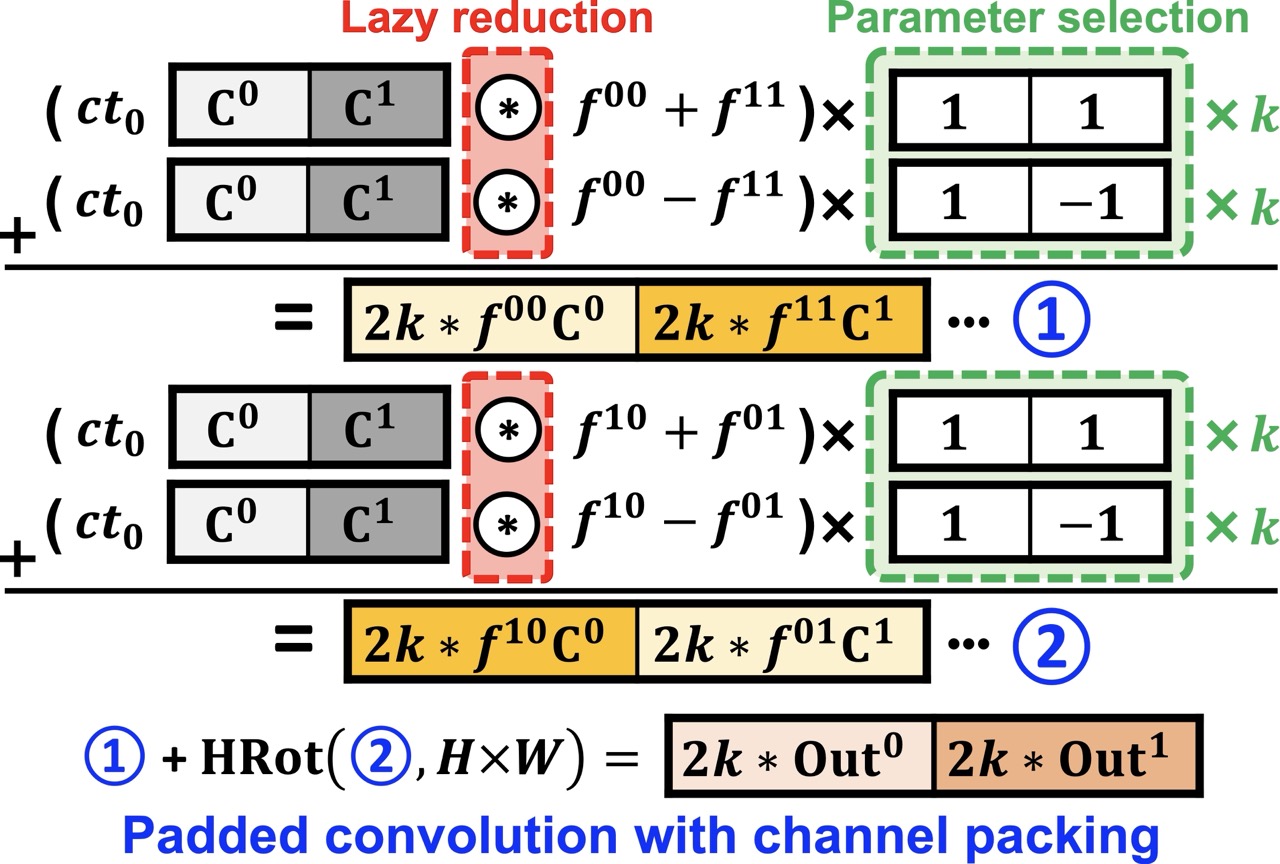}
\vspace{0em}
\caption{Proposed convolution with multiple input and output channels enabling channel packing.}
\vspace{-1em}
\label{proposed}
\end{figure}

Let $w$ be a plaintext whose slots are filled with two kernel elements, say $[f_0, f_1]$, where the first $n/2$ slots are filled with $f_0$ and the rest half with $f_1$,
and $\llbracket m_0\rrbracket$ be a ciphertext containing two input channels.
To obtain partial sums packed in a ciphertext, $\textbf{PMult}(\llbracket m_0\rrbracket, w)$ needs to be performed, but this needs $[f_0, f_1]$ to be stored as the plaintext $w$, requiring the same amount of model storage capacity as the conventional convolution.
Instead, we note that, $[f_0, f_1]^\intercal$ can be expressed as  
\begin{equation}
\begin{bmatrix}
    f_{0} \\
    f_{1}
    \end{bmatrix}
    =
    \frac{1}{2}
    \left[ \begin{array}{cc}
                        1 & 1 \\
                        1 & -1 
            \end{array} 	
    \right] 
    \times
    \begin{bmatrix}
    f_{0}+f_1 \\
    f_{0}-f_1
    \end{bmatrix},
\label{hadamard2}
\end{equation}
and the plaintext whose slots are filled with only 1 is simply the constant 1. 
This implies that $\textbf{PMult}(\llbracket m_0\rrbracket,w)$ is equivalent to $\textbf{CMult}(\llbracket m_0\rrbracket, \frac{f_0+f_1}{2})+\textbf{PMult}(\textbf{CMult}(\llbracket m_0\rrbracket,\frac{f_0-f_1}{2}),w')$, where $w'$ is a plaintext whose slots are filled with $[1, -1]$.
Note that {\bf PMult} with a different $w$ (e.g., packing $[f_2, f_3]$) makes only the {\bf CMult} scalars different and $w'$ remain the same.
Using this observation, rather than storing huge number of plaintexts packing kernel elements as in conventional convolutions, 
we can store only the integers as many as kernel elements and a single polynomial whose slots are filled with $[1, -1]$, 
providing the low storage requirement benefit as padded convolution.
For instance, if $n$=2048 and the ciphertext modulus $q$ is 60 bits, for a convolution layer with the kernel dimension $2\times 3\times3$,
the proposed method requires only 16.1\,KB for all the weights, whereas 288\,KB is needed for the conventional method.
Note that a larger amount of saving can be achieved as the number of kernel elements increases.

This approach can be easily generalized to the case when the number of channels packed in a ciphertext, say $c_n$, becomes larger than 2.
Since $n$ is a power of 2, for simplicity we assume that the zero-padded input channel width is a power of 2 so $c_n$ also can become a power of 2.
Then by choosing a $c_n \times c_n$ matrix whose columns are orthogonal each other, 
the same process as above can be done.
One matrix that satisfies such property is the Walsh-Hadamard matrix $\mathbf{H_{c_n}}$ whose elements are binary, i.e., +1 or -1.
For instance, if $c_n=4$, i.e., the number of channels packed in a ciphertext is 4, 
we use the basis \{[1, 1, 1, 1], [1, -1, 1, -1], [1, 1, -1, -1], [1, -1, -1, 1]\} and follow the same process.

The complete process of Hyena is described in Fig.~\ref{proposed}. 
Here, we assume the same convolution layer as in Fig.~\ref{conventional}.
The partial sums for each output channel (indicated as \textcircled{1} and \textcircled{2} in Fig.~\ref{proposed}) are computed using the aforementioned technique.
Then, to obtain the final convolution output, {\bf HRot} is utilized to align the partial sums before addition.
In terms of required model storage capacity and communication cost,
the proposed convolution has 1) as low storage as padded convolution by storing kernel elements in integers, and 2) as low communication cost as conventional packed convolution by enabling output channel packing.
However, the proposed convolution should perform $c_n$ {\bf CMult} and {\bf PMult} operations rather than a single {\bf PMult},
which implies that as $c_n$ gets larger, the amount of computation also increases, while the communication cost decreases proportionally. 
We note that majority of layers in deep CNNs for ImageNet have large input feature maps ($H,W \geq 32$), 
so optimizing the proposed convolution latency for $c_n \leq 2$ provides large end-to-end latency reduction as will be discussed in Section~\ref{sec:results}.
Hence, in the following, we describe how the proposed convolution can be accelerated when $c_n=2$.\footnote{The same approach can be also applied when $c_n>2$.}

\subsection{Latency Optimization}
\subsubsection{Optimal Parameter Selection}

The main idea behind this optimization is to find the optimal encryption parameters that minimize the noise growth during the proposed convolution.
This noise budget saving, in turn, allows us to choose a larger decomposition base for computationally costly {\bf HRot} (see Section~\ref{sec:background}), thereby reducing the overall amount of computation.

For optimal parameter selection, we should understand how the encryption parameters affect noise in ciphertext during convolution.
As described in Section~\ref{sec:background}, {\bf PMult} boosts noise by a factor of $np/2$ in the worst case, while {\bf CMult} with $f_0$ does by a factor of only $f_0$, which is much less than $p$.
Moreover, in Hyena, {\bf PMult} is always performed with the fixed plaintext polynomial packing $[1,-1]$ in its slots. 
We note that such polynomial has only two non-zero terms and all the other coefficients are zero.
Furthermore, the two non-zero coefficients in this sparse polynomial are identical, and their values are determined by $p$ and $n$.
If we denote the non-zero coefficient as $h_{p,n}$, 
then {\bf PMult} with the plaintext containing $[1,-1]$ boosts noise by a factor of $2h_{p,n}$ only. 
Hence, performing {\bf PMult} with a plaintext packing $[f_0, f_1]$ in Hyena increases noise by a factor of $(f_0+f_1 + 2(f_0-f_1)h_{p,n})/2$, as opposed to $np/2$ in conventional {\bf PMult}.
For instance, if the same encryption parameters as in Gazelle~\cite{juvekar2018gazelle} are used, i.e., $n$=2048 and $p$=307201 (19-bit prime), then $h_{p,n}$=84248 (17 bits).
Since $f_0$ and $f_1$ are the kernel elements typically less than 8 bits, 
{\bf PMult} in Hyena increases noise by at most 25 bits, while the conventional method does by 29 bits.
Due to the less noise growth, Hyena can choose a larger decomposition base for {\bf HRot}, leading to lower computational overhead.

Going one step further, if we choose the encryption parameters ($p, n$) such that $h_{p,n}$ can be minimized,
noise growth can be reduced more, resulting in a further decrease in computation.
To this end, we note that the plaintext containing $[k,-k]$ ($k \in \mathbb{Z}$) also has two non-zero coefficients, which are $kh_{p,n}\ (\textrm{mod}\ p)$.\footnote{{\bf PMult} with a plaintext containing $[k,-k]$ does not change the inference result because all the convolution outputs are scaled by the same amount $k$.}
Since noise growth in {\bf PMult} is proportional to the plaintext coefficient values, 
for given $p$ and $n$, noise can be minimized by finding an integer $k$ to minimize $kh_{p,n}\ (\textrm{mod}\ p)$.
Based on this insight, our parameter selection procedure is done as follows:
1) To enable SIMD packing, find $(n,p,q)$ such that $p \equiv 1 \ (\textrm{mod}\ 2n)$ and  $q \equiv 1 \ (\textrm{mod}\ 2n)$.
2) For each $(n,p)$ obtained from 1), compute $h_{p,n}$, the non-zero coefficient of the plaintext polynomial packing $[1,-1]$. 
3) Find an integer $k$ such that $kh_{p,n}\ (\textrm{mod}\ p)$ is minimized.
For instance, from this procedure, we can obtain $k$=11 when $n$=2048 and $p$=307201, and {\bf PMult} with a plaintext containing $[-11, 11]$ increases noise by at most 21 bits only.
\ignore{It is possible to precompute both $h_{p,n}$ and $k$ for all utilized $n$ and $p$.}
Employing this technique saves additional noise budget and allows Hyena to choose an even larger decomposition base for {\bf HRot}, thereby reducing the computation latency.

\begin{table}
\setlength{\tabcolsep}{4pt}
\centering
\vspace{0em}
\caption{Runtime ($\mu$s) for HE operations with polynomial degrees ($n$) of 2048. 
HAdd between 128-bit coefficient ciphertexts without output reduction is denoted as $\text{HAdd}^{128}$.
CMult without reduction is denoted as $\text{CMult}^{-}$.}
\label{operation}
{\small
\begin{tabular*}{\columnwidth}{@{\extracolsep{\fill}}C{1cm}|cc|ccC{1.5cm}}
\toprule[1.5pt]
\multirow{1}{*}{\textbf{$n$}} & \multirow{1}{*}{\textbf{HAdd}} & \multirow{1}{*}{\textbf{HAdd$^{128}$}} & \multirow{1}{*}{\textbf{CMult}} & \multirow{1}{*}{\textbf{CMult$^{-}$}} & \multirow{1}{*}{\textbf{Reduction}}\\
\midrule[0.2pt]
\midrule[0.2pt]
2048 & 4 & 7 & 19 & 3 & 16.5 \\
\bottomrule[1.5pt]
\end{tabular*}
}
\vspace{-1em}
\end{table}

\begin{figure}[t]
\centering
\includegraphics[width=\columnwidth]{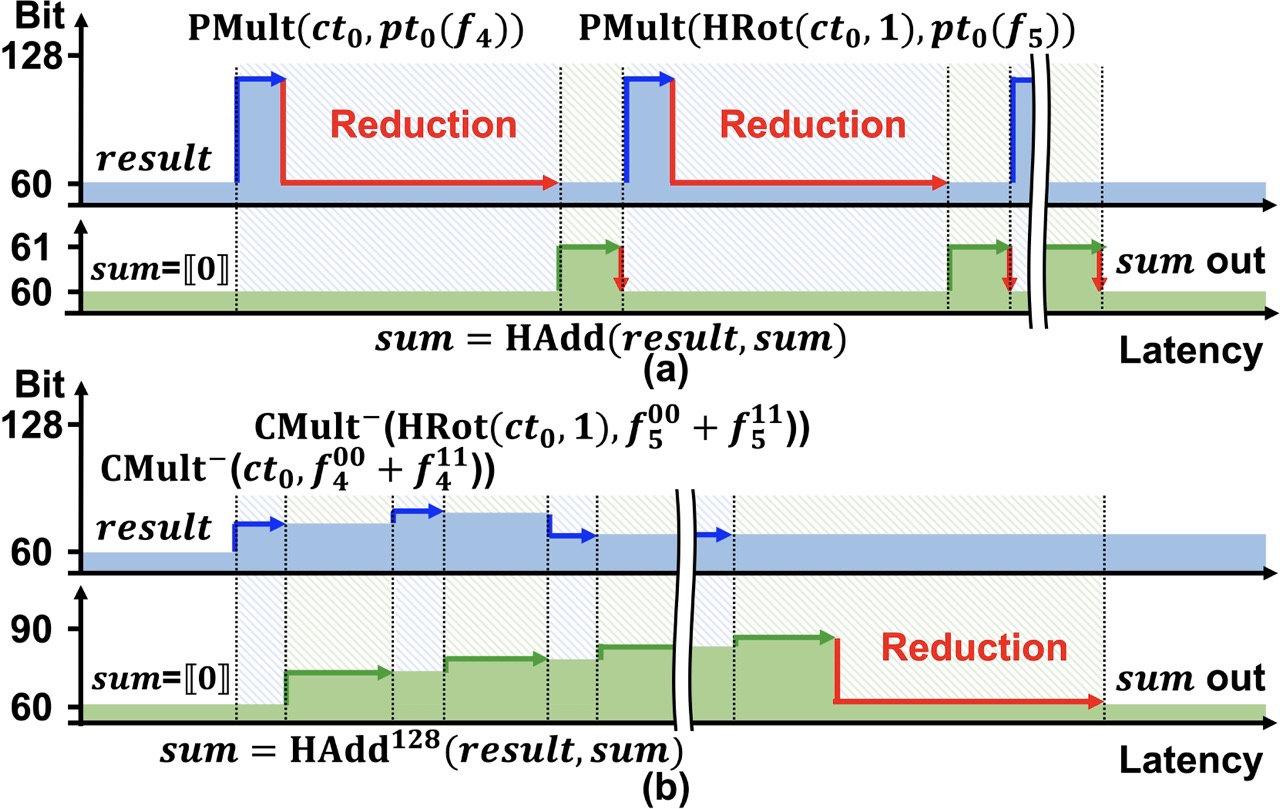}
\vspace{-2em}
\caption{
Bit width of ciphertext coefficients during HE convolution. $result$ is to store each multiplication output, and $sum$ is to accumulate the partial sums to obtain the final convolution result. {$\llbracket \textbf{0}\rrbracket$ denotes an empty ciphertext.} Blue areas represent HE multiplication operations ($\textbf{PMult}$ or $\textbf{CMult}^{-}$), and green areas represent HE addition operations ($\textbf{HAdd}$ or $\textbf{HAdd}^{128}$): (a) conventional convolution, and (b) proposed convolution with lazy reduction.}
\vspace{-1em}
\label{fig:lazy}
\end{figure}

\begin{figure*}[t]
\centering
\includegraphics[width=1.7\columnwidth]{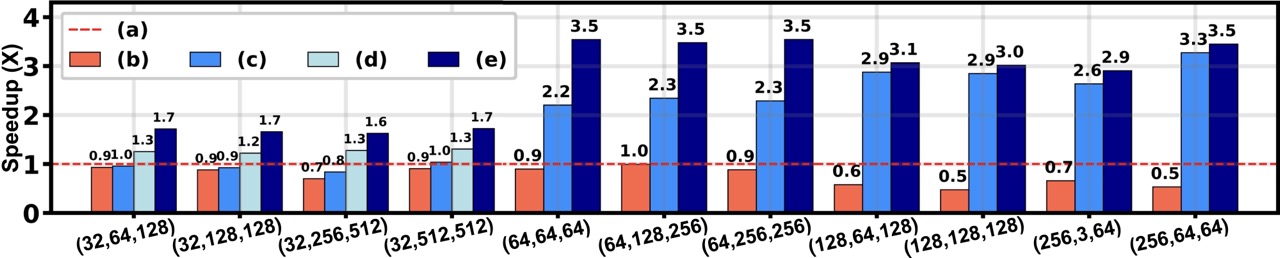}
\vspace{-1em}
\caption{Normalized convolution latency where the tuple represents (input channel width, input channel count, output channel count): (a) conventional convolution (baseline), (b) conventional convolution without hoisting technique, (c) proposed convolution without latency optimization, (d) proposed convolution with optimal parameter selection only, and (e) proposed convolution with all optimization techniques.}
\vspace{-1em}
\label{fig:latency}
\end{figure*}

\subsubsection{Lazy Reduction}

Due to modular arithmetic operations in HE, every addition or multiplication is followed by reduction. 
In this work, we use 128-bit data type to store the temporary results produced during convolution.
As long as the temporary value size remains below 128 bits, we can utilize lazy reduction (i.e., skipping intermediate reductions) to reduce computation. 
In the convolution process illustrated in Fig.~\ref{proposed}, a ciphertext with 60-bit coefficients undergoes multiplication by scalars and subsequent addition. 
Since the multiplied scalars are less than the plaintext modulus $p$, 
the product after {\bf CMult} is less than $\log_2 pq$ bits (79 bits in this work).
Then, even with hundreds of channels, adding those products stays below 120 bits.
\ignore{Thus, rather than conducting reduction after each multiplication and addition \fixme{as Fig.~\ref{fig:lazy}(a)}, we only perform it once: after summing to obtain the intermediate partial sums, just before {\bf PMult} with a plaintext containing [1, -1].}
Thus, unlike in Fig.~\ref{fig:lazy}(a) where reductions are conducted after each multiplication and addition\footnote{Reduction in modular addition can be done by comparison and subtraction, so it is fast as shown in Fig.~\ref{fig:lazy}(a).}, we only perform it once. 
After summing $result$'s to obtain the intermediate partial sums, reduction is executed just before {\bf PMult} with a plaintext containing $[k, -k]$.
128-bit data type is a non-native data type for 64-bit machines, and 128-bit addition is slower than 64-bit addition, but this latency increase is much smaller than what we can save from lazy reduction. 
As highlighted in Table~\ref{operation}, reduction is the most time-consuming operator.
By employing the proposed lazy reduction as Fig.~\ref{fig:lazy}(b), computational overhead incurred by reduction, typically composed of several integer multiplication and addition, can be greatly saved.
\ignore{In convolution calculations involving addition and multiplication, addition isn't the dominating factor, as evidenced by Table~\ref{operation}.
\ignore{The multiplication latency, which is the bottleneck, is significant reduced because it passes the reduction.
This optimized lazy reduction minimizes the computational load, addressing the main delay in homomorphic multiplication caused by intensive integer division.}
The lazy reduction minimizes the computational load, addressing the main delay in homomorphic multiplication caused by intensive integer division.}

\section{Experimental Results}
\label{sec:results}

\subsection{Evaluation Setup}
Hyena was implemented in C++ with the SEAL library~\cite{seal} and
tested on a 1600\,MHz Intel Xeon Gold processor with 256\,GB of RAM and 1536/24576/36608\,KB of L1/L2/L3 cache memory. 
The model parameters for the experiments are obtained from VGG-16, ResNet-20, and MobileNetV1 for ImageNet.
$n$ is chosen as either 2048 or 4096 based on the input size. 
Specifically, for $32\times 32$-sized inputs, we use 2048, while for larger inputs, we opt for 4096. 
This choice makes $c_n \leq 2$ for majority of the layers in the tested models.
We set the same encryption parameters as Gazelle~\cite{juvekar2018gazelle} guaranteeing 128-bit security level ($p$ as a 19-bit prime and $q$ as a 60-bit prime).

\subsection{Convolution Evaluation}
In order to quantify the effectiveness of the proposed optimization techniques described in Section~\ref{sec:proposed},
we compare the convolution latency in Fig.~\ref{fig:latency}.
Fig.~\ref{fig:latency}(b) is the conventional SIMD-based convolution implemented without the hoisting technique, achieving 0.47-0.96$\times$ slower than that with hoisting.
We can observe that the naive implementation of the proposed convolution without any latency optimization (denoted as (c) in Fig.~\ref{fig:latency}) when $c_n$=2 is 0.83-0.96$\times$ slower than the conventional convolution (denoted as (a)) because of the increased number of operations as described in Section~\ref{sec:proposed}.
However, by minimizing the noise through optimal parameter selection, we can choose a larger decomposition base for {\bf HRot} and reduce the amount of computation.
This accelerates the naive implementation (c), thereby achieving 1.12-1.3$\times$ speedup compared to the conventional convolution, as seen in Fig.~\ref{fig:latency}(d).
For inputs larger than $32\times 32$, since $n$ is either 2048 or 4096, packing is not required, i.e., $c_n$=1.
In this case, we can also leverage a large decomposition base for {\bf HRot} thanks to small noise increase after {\bf CMult}.
Consequently, the proposed convolution without lazy reduction (c) results in a speed-up of 2.1-3.3$\times$ over the conventional method when $c_n$=1.
Incorporating lazy reduction, the proposed convolution with optimal performance (e) becomes 1.63-3.84$\times$ faster than the conventional method.

\begin{table}
\centering
\caption{Comparison of required storage capacity and communication costs\ignore{input ciphertext size} between conventional convolution and Hyena with filter dimensions $f_w \times f_h$ of $3\times 3$. \ignore{The input ciphertext, labeled as "Input ct.," represents the encrypted data size sent from the client to the server.}}
\label{summary}
{\small
\begin{tabular*}{\columnwidth}{@{\extracolsep{\fill}}ccccc}
\toprule[1.5pt]
\multicolumn{1}{c}{\multirow{2}{*}[-0.65ex]{\parbox{1.5cm}{\centering {\textit{\textbf{Input}\\\scriptsize{(H=W, in, out)}}}}}} & \multicolumn{1}{c}{\multirow{2}{*}[-0.65ex]{\parbox{1.25cm}{\centering \textbf{\textit{CONV}}}}} & \multicolumn{2}{c}{\textbf{\textit{Storage (MB)}}} & \multicolumn{1}{c}{\multirow{2}{*}[-0.65ex]{\parbox{1.22cm}{\centering \textbf{\textit{Commu. (MB)}}}}}\\
\cmidrule{3-4}
& & \textbf{$\text{Key}_{\text{}}$} & \textbf{Model}\\
\midrule[0.2pt]
\midrule[0.2pt]
\multirow{2}{*}{(32, 64, 128)} & \scriptsize{Conventional} & {1.5} & {1153} & {3}\\
& \scriptsize{\textbf{Proposed}} & \textbf{{2.5} ({0.6}$\times$)} & \textbf{{0.56} ({2048}$\times$)} & \textbf{{2} ({1.5}$\times$)}\\[0.5mm]
\multirow{2}{*}{(32, 128, 128)} & \scriptsize{Conventional} & {1.5} & {2306} & {6}\\
& \scriptsize{\textbf{Proposed}} & \textbf{{2.5} ({0.6}$\times$)} & \textbf{{1.13} ({2048}$\times$)} & \textbf{{4} ({1.5}$\times$)}\\[0.5mm]
\multirow{2}{*}{(32, 256, 512)} & \scriptsize{Conventional} & {2} & {18446} & {12}\\
& \scriptsize{\textbf{Proposed}} & \textbf{{2.5} ({0.8}$\times$)} & \textbf{{9} ({2048}$\times$)} & \textbf{{8} ({1.5}$\times$)}\\[0.5mm]
\multirow{2}{*}{(32, 512, 512)} & \scriptsize{Conventional} & {2} & {36891} & {24}\\
& \scriptsize{\textbf{Proposed}} & \textbf{{2.5} ({0.8}$\times$)} & \textbf{{18} ({2048}$\times$)} & \textbf{{16} ({1.5}$\times$)}\\[0.5mm]
\multirow{2}{*}{(64, 64, 64)} & \scriptsize{Conventional} & {1.5} & {2305} & {12}\\
& \scriptsize{\textbf{Proposed}} & \textbf{{2} ({0.75}$\times$)} & \textbf{{0.28} ({8192}$\times$)} & \textbf{{8} ({1.5}$\times$)}\\[0.5mm]
\multirow{2}{*}{(64, 128, 256)} & \scriptsize{Conventional} & {4} & {18439} & {24}\\
& \scriptsize{\textbf{Proposed}} & \textbf{{2} ({2}$\times$)} & \textbf{{2.25} ({8192}$\times$)} & \textbf{{16} ({1.5}$\times$)}\\[0.5mm]
\multirow{2}{*}{(64, 256, 256)} & \scriptsize{Conventional} & {4} & {36878} & {48}\\
& \scriptsize{\textbf{Proposed}} & \textbf{{2} ({2}$\times$)} & \textbf{{4.5} ({8192}$\times$)} & \textbf{{32} ({1.5}$\times$)}\\[0.5mm]
\multirow{2}{*}{(128, 64, 128)} & \scriptsize{Conventional} & {2} & {2306} & {48}\\
& \scriptsize{\textbf{Proposed}} & \textbf{{1} ({2}$\times$)} & \textbf{{0.56} ({4096}$\times$)} & \textbf{{32} ({1.5}$\times$)}\\[0.5mm]
\multirow{2}{*}{(128, 128, 128)} & \scriptsize{Conventional} & {2} & {4611} & {96}\\
& \scriptsize{\textbf{Proposed}} & \textbf{{1} ({2}$\times$)} & \textbf{{1.13} ({4096}$\times$)} & \textbf{{64} ({1.5}$\times$)}\\[0.5mm]
\multirow{2}{*}{(256, 3, 64)} & \scriptsize{Conventional} & {2} & {54.04} & {9}\\
& \scriptsize{\textbf{Proposed}} & \textbf{{1} ({2}$\times$)} & \textbf{{0.01} ({4096}$\times$)} & \textbf{{6} ({1.5}$\times$)}\\[0.5mm]
\multirow{2}{*}{(256, 64, 64)} & \scriptsize{Conventional} & {2} & {1153} & {192}\\
& \scriptsize{\textbf{Proposed}} & \textbf{{1} ({2}$\times$)} & \textbf{{0.28} ({4096}$\times$)} & \textbf{{128} ({1.5}$\times$)}\\
\bottomrule[1.5pt]
\end{tabular*}
}
\vspace{-1em}
\end{table}

\begin{figure}[t]
\centering
\includegraphics[width=0.9\columnwidth]{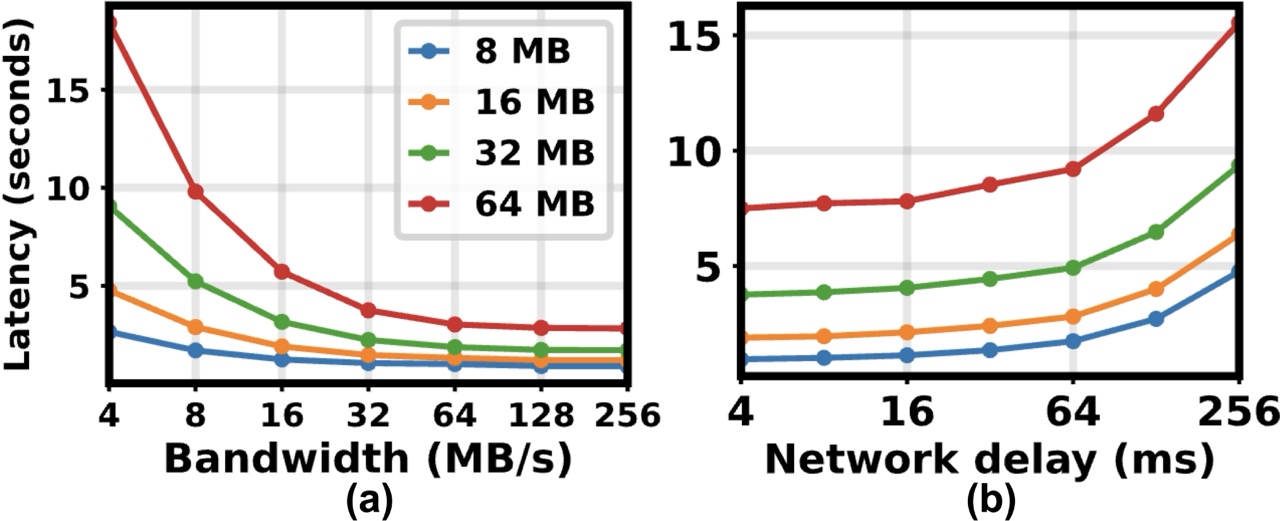}
\vspace{-1em}
\caption{Communication latency with various (a) network bandwidth and (b) network delay.}
\vspace{-1em}
\label{fig:bandwidth}
\end{figure}


\begin{table}
\begin{threeparttable}
\centering
\caption{Comparison of the latency, communication costs, and \ignore{memory}{storage} usage between Hyena and the state-of-the-art SIMD-based convolution~\cite{sci,rathee2020cryptflow2}. $f$ denotes $f_w$ and $f_h$, which are the same.}
\label{communication}
{\small
\begin{tabular*}{\columnwidth}{@{\extracolsep{\fill}}ccccc}
\toprule[1.5pt]
\multicolumn{1}{c}{\multirow{2}{*}[-0ex]{\parbox{1.cm}{\centering \textbf{\textit{CONV}}}}} & \multicolumn{1}{c}{\multirow{2}{*}[-0.ex]{\parbox{2.cm}{\centering \textbf{\textit{Input\\{(H=W, in, out, f)}}}}}} & \multicolumn{1}{c}{\multirow{2}{*}[-0.ex]{\parbox{1.1cm}{\centering \textbf{\textit{Latency\footnotemark[1] (s)}}}}} & \multicolumn{1}{c}{\multirow{2}{*}[-0.ex]{\parbox{1.1cm}{\centering \textbf{\textit{Commu. (MB)}}}}} & \multicolumn{1}{c}{\multirow{2}{*}[-0.ex]{\parbox{1.1cm}{\centering \textbf{\textit{Storage (MB)}}}}}\\
&&&& \\
\midrule[0.2pt]
\midrule[0.2pt]
\multicolumn{1}{c}{\multirow{4}{*}[-0ex]{\parbox{1.3cm}{\centering {\cite{rathee2020cryptflow2,sci}}}}} 
& (224, 3, 64, 3)  & {7.06} & {76.02} & {55.54} \\ 
& (56, 64, 256, 1)  & {8.21} & {28.01} & {1024.8} \\ 
& (56, 256, 64, 1)  & {7.41} & {52.02} & {1024.8} \\ 
& (28, 192, 3)\footnotemark[2]  & {12.01} & {12.07} & {65.05} \\ 
\cmidrule{1-5}
\multicolumn{1}{c}{\multirow{4}{*}[-0ex]{\parbox{1.3cm}{\centering {Hyena}}}} 
& (224, 3, 64, 3)  & {7.3} & {67} & {0.76} \\ 
& (56, 64, 256, 1)  & {2.16} & {20} & {0.13} \\ 
& (56, 256, 64, 1)  & {1.87} & {20} & {0.13} \\ 
& (28, 192, 3)\footnotemark[2]  & {0.14} & {{5.06}} & {2.64} \\
\bottomrule[1.5pt]
\end{tabular*}
}
\begin{tablenotes}
\footnotesize
\item [1] Latency includes encryption, convolution, decryption runtime and communication latency with a network bandwidth of 384\,MB/s and a delay of 0.3\,ms.
\item [2] This tuple represents the input features, channels, and kernel size for the depth-wise convolution.
\end{tablenotes}
\end{threeparttable}
\vspace{-1em}
\end{table}

{Table~\ref{summary} shows the convolution layers contained in VGG-16, ResNet-20 and MobileNetV1. 
\ignore{In addition to key generation time, Gazelle also requires offline time to convert all filters to NTT plaintext for each convolution. If these filter plaintexts are not stored in advance, the offline phase must be performed before each convolution, which significantly increases the overall running time.
\ignore{Therefore, they usually use huge memory space to store and reuse them.}
To mitigate this issue, Gazelle often employs a large memory space to store and reuse the filter plaintexts.
In contrast, the proposed convolution eliminates the need for converting filters to plaintexts, as the Hadamard transform can be quickly applied to the filter, resulting in the elimination of offline time for filter preparation.
During the online phase, the padded convolution is performed using the prepared Hadamard encoded scalar.
As previously demonstrated, the amount of noise increase resulting from multiplying by a scalar is proportional to the scalar value.
Therefore, the accumulated noise is smaller than that of multiplying by a 64-bit coefficient, as in the case of Gazelle.
This allows us to use a larger decomposition base. The larger decomposition of data plaintexts provides an additional noise budget, enabling us to set the plaintext decomposition base to 20, thereby avoiding the doubling of latency, memory usage, and communication cost.}
To process such convolution layers, cloud requires key-switching public keys for {\bf HRot} and the input ciphertexts, as well as the model weights. 
Preparing the weights in the proper format (i.e., in plaintexts for conventional convolution and in integers for Hyena) and generating the public keys are typically done during offline phase to reduce online inference latency, and the generated elements should be stored in cloud.
Furthermore, in case of the conventional HE convolution, 
to manage noise due to {\bf PMult} the plaintext decomposition technique described in \cite{juvekar2018gazelle} should be employed, which increases the amount of computation and the size of the input ciphertexts and weight plaintexts.
For the convolution layers in Table~\ref{summary}, this plaintext decomposition doubles the size of the input ciphertext and weight plaintexts, which necessitates a large memory space.
In contrast, Hyena obviates the need to perform plaintext decomposition and to store the weights in the plaintext format, thereby greatly saving storage.
\ignore{In summary of Table~\ref{summary}, compared to conventional convolution, the proposed has benefits of 0.88-7.29$\times$ in offline runtime, 1.63-3.8$\times$ in online runtime, 0.6-2$\times$ in permutation key storage, 2048-8192$\times$ in model storage, and 2$\times$ in client-to-server communication costs. 
}
Required storage capacity is reduced by 0.6-2$\times$ for the public keys and dramatically by 2048-8192$\times$ for the model weights.
Additionally, since halving the input ciphertext size, Hyena results in a 1.5$\times$ reduction in total communication costs.

\begin{figure}[t]
\centering
\includegraphics[width=\columnwidth]{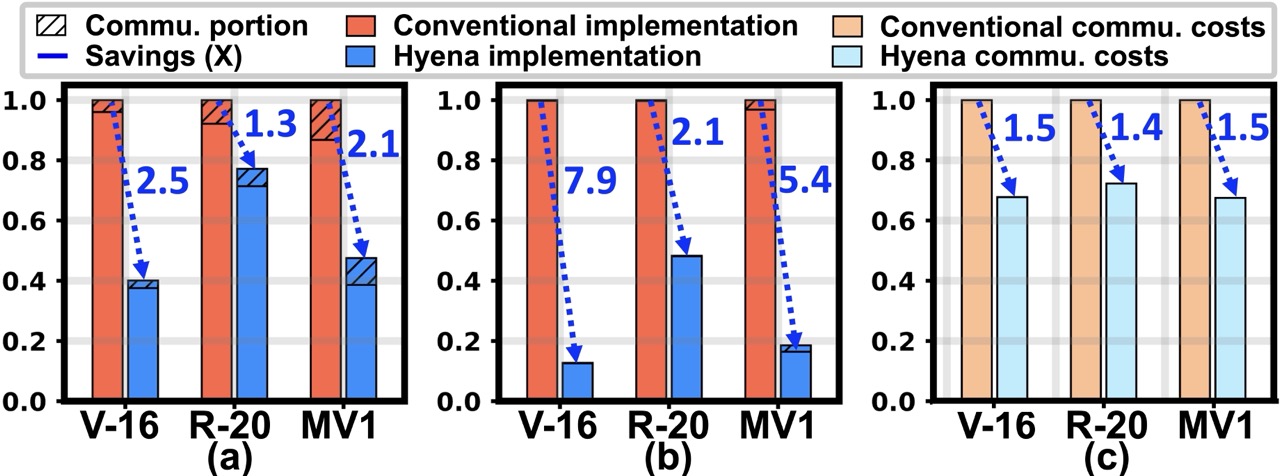}
\vspace{-2em}
\caption{End-to-end performance comparison for VGG (V)-16, ResNet (R)-20, and MobileNetV1 (MV1): (a) latency, (b) memory {usage} (including keys, model weights, and input/output ciphertexts size), and (c) communication costs\ignore{ for linear layers}. All values are normalized to conventional implementation.}
\vspace{-1em}
\label{fig:total}
\end{figure}

\ignore{We employ the MPC framework, like Gazelle~\cite{juvekar2018gazelle} and Cheetah~\cite{reagen2020cheetah}, to structure the entire network.
The server executes linear layers while both client and server collaborate on non-linear layers.
During non-linear computation, they communicate using a ``garbled circuit" that supports all boolean operations, enabling lossless execution of the ReLU function at the cost of communication. 
The key insight here is that the entire process retains accuracy as long as the linear phase remains lossless.
Given that the proposed method preserves all filter values, conventional and proposed implementations avoid loss of accuracy.}

\subsection{Communication Latency Evaluation}


Using Linux Traffic Control as~\cite{xu2023falcon}, communication latency sensitivity to network bandwidth (MB/s) and network delay (ms) is demonstrated in Fig.~\ref{fig:bandwidth}. 
In  Fig.~\ref{fig:bandwidth}(a), communication latency is measured with a fixed network delay of 50\,ms, and in Fig.~\ref{fig:bandwidth}(b) the bandwidth is fixed at 9\,MB/s as in~\cite{xu2023falcon,mohassel2017secureml}.
As communication costs (i.e., input and output ciphertext size) increase from 8\,MB to 64\,MB, communication latency proportionally increases.
For the convolution layer with ($128, 128, 128, 3$) incurring a communication cost of 64\,MB, the latency reaches 18 seconds at a network bandwidth of 4\,MB/s and delay of 50\,ms.

While the runtime of HE convolution usually exceeds the communication latency in end-to-end implementations, the latter can significantly impact the overall latency depending on the network condition.
Latency degradation becomes more pronounced with each doubling of communication costs.
Therefore, saving communication costs by enabling output channel packing in a ciphertext as in Hyena is essential to prevent unpredictable latency.

\subsection{End-to-end Evaluation}
\ignore{The optimized network performance is demonstrated through its reduced latency, memory usage, and communication cost compared to Gazelle in both the offline and online phases. }
We incorporate Hyena into VGG-16 by replacing the first ten of thirteen convolution layers (CONV1 to 10). 
Similarly, for ResNet-20 and MobileNetV1, we apply Hyena to the first nine of seventeen (CONV1 to 9) and the first 24 of 27 convolution layers (CONV1 to 24), respectively.
The conventional convolution method is employed for the rest of the layers with the input sizes of 16$\times$16 or 8$\times$8.
Performance is evaluated with the network bandwidth of 9\,MB/s and delay of 50\,ms.

\ignore{Fig.~\ref{fig:total} summarizes the end-to-end performance improvement for VGG-16, ResNet-20, and MobileNetV1 by utilizing the proposed convolution.
Fig.~\ref{fig:total}(a) compares the normalized total linear layer computation latency between the proposed and conventional networks.
The conventional implementation reports a latency of 1971 seconds for VGG-16 and 147 seconds for ResNet-20.
For ResNet-18, it stands at 53 seconds, which, thanks to the hoisting technique, is faster than the 17.7 minutes observed in prior work~\cite{garimella2023characterizing}.}
Fig.~\ref{fig:total} summarizes the end-to-end performance improvement by employing Hyena.
\ignore{Fig.~\ref{fig:total}(a) compares the normalized full network latency, with the nonlinear layer latency occupying 12-39\,\% in each network.}
Fig.~\ref{fig:total}(a) compares the normalized full network latency, while the communication latency occupies 3-13\,\% in each network.
For the linear layers, the conventional implementation reports a latency of 1971 seconds for VGG-16 and 147 seconds for ResNet-20.
For MobileNetV1, it stands at 1022 seconds.
In contrast, Hyena achieves a latency of 771 seconds for VGG-16, 114 seconds for ResNet-20, and 455 seconds for MobileNetV1 showing a speedup of \ignore{1.30-2.56}1.3-2.6$\times$. 
The communication latency improves by 1.4-1.6$\times$, decreasing from 82, 13, and 156 seconds to 52, 9, and 105 seconds, respectively.
Hence, Hyena demonstrate a 1.3-2.5$\times$ reduction in end-to-end latency.

In Fig.~\ref{fig:total}(b), memory usage required for storing key-switching public keys and model weights, and input/output ciphertexts are illustrated.
VGG-16 shows the most significant reduction, with its model size dropping from 215\,GB to 27\,GB, an 8-fold decrease. 
Hyena, specifically for CONV1-10, contribute to a 3307$\times$ reduction. 
ResNet-20's model is halved, dropping from 31\,GB to 16\,GB, with the optimized CONV1-9 sections achieving a 3416$\times$ reduction.
\ignore{Similarly, ResNet-18 is cut down by half, from 14\,GB to 7\,GB, thanks to CONV1-5, which undergoes a 3612$\times$ decrease. 
Considering the key size, the proposed networks reduce memory storage between 2.1 to 7.9$\times$.}
{Similarly, MobileNetV1 is reduced from 36\,GB to 6\,GB, thanks to CONV1-24, which undergoes a 2482$\times$ decrease. 
Including the key size and input/output ciphertexts, the proposed networks reduce the total memory storage by 2.1-7.9$\times$.}
Fig.~\ref{fig:total}(c) shows the communication cost reduction for linear layers. 
Hyena helps reduce the total communication cost by 1.4-1.5$\times$, decreasing from 543\,MB, 93.8\,MB, and 1203\,MB to 368\,MB, 67.8\,MB, and 812\,MB for VGG-16, ResNet-20, and MobileNetV1, respectively.

\ignore{As demonstrated in Table~\ref{summary}, eliminating plaintext decomposition results in a twofold reduction of input ciphertext, leading to a saving of 1.7-2.0$\times$ across the three networks.}

\ignore{There is no decrease in accuracy; it is the same as when using Gazelle. 
Since MPC is assumed, the existing neural network structure is not modified. 
Since the encoded filter value can be recovered without approximation, convolution is carried out using the original filter value, meaning that the proposed structure has no impact on accuracy.}

{Lastly, we compare the latency, communication cost, and {storage} usage with another state-of-the-art SIMD-packing based convolution CrypTFlow2~\cite{rathee2020cryptflow2} in Table~\ref{communication}. 
The output channel packing technique of Hyena allows the slots to be fully utilized without wastage, which leads to a 1.13-2.6$\times$ reduction in communication cost for large input sizes (>32, $c_n$=1). 
Additionally, for the 28$\times$28 size with $c_n$=2, the communication cost is reduced by 4.2$\times$.}
{Thanks to the proposed optimization, latency is also reduced by {0.97-86$\times$, and storage savings range from 24.6-7883$\times$.}}

\begin{table}
\begin{threeparttable}
\centering
\caption{Comparison of the latency, communication costs, and \ignore{memory}{storage} usage between Hyena and the state-of-the-art non-SIMD-based convolution~\cite{huang2022cheetah}.}
\label{tab:cheetah}
{\small
\begin{tabular*}{\columnwidth}{@{\extracolsep{\fill}}@{}c@{}c@{}ccc}
\toprule[1.5pt]
\multicolumn{1}{c}{\multirow{2}{*}[-0ex]{\parbox{1.cm}{\centering \textbf{\textit{CONV}}}}} & \multicolumn{1}{c}{\multirow{2}{*}[-0.ex]{\parbox{2.cm}{\centering \textbf{\textit{Input\\{(H=W, in, out, f)}}}}}} & \multicolumn{1}{c}{\multirow{2}{*}[-0.ex]{\parbox{1.1cm}{\centering \textbf{\textit{Latency\footnotemark[1] (s)}}}}} & \multicolumn{1}{c}{\multirow{2}{*}[-0.ex]{\parbox{1.1cm}{\centering \textbf{\textit{Commu. (MB)}}}}} & \multicolumn{1}{c}{\multirow{2}{*}[-0.ex]{\parbox{1.1cm}{\centering \textbf{\textit{Storage (MB)}}}}}\\
&&& \\
\midrule[0.2pt]
\midrule[0.2pt]
\multicolumn{1}{c}{\multirow{4}{*}[-0ex]{\parbox{1.3cm}{\centering {\cite{huang2022cheetah}}}}} 
& (28, 192, 192, 1)  & {9.84} & {8.7} & {243}\\
& (28, 192, 3)\footnotemark[2]  & {1.26} & {8.7} & {3}\\
& (14, 384, 384, 1)  & {15.28} & {12.7} & {288}\\
& (14, 384, 3)\footnotemark[2]  & {1.95} & {12.7} & {6}\\
\cmidrule{1-5}
\multicolumn{1}{c}{\multirow{4}{*}[-0ex]{\parbox{1.3cm}{\centering {Hyena}}}} 
& (28, 192, 192, 1)  & {3.89} & {5.06} & {0.34}\\ 
& (28, 192, 3)\footnotemark[2]  & {0.86} & {5.06} & {2.64}\\ 
& (14, 384, 384, 1)  & {10.12} & {7.5} & {1.18}\\ 
& (14, 384, 3)\footnotemark[2]  & {1.31} & {7.5} & {2.65}\\ 
\bottomrule[1.5pt]
\end{tabular*}
}
\begin{tablenotes}
\footnotesize
\item [1] Latency includes encryption, convolution, decryption runtime and communication latency with a network bandwidth of 9\,MB/s and a delay of 10\,ms.
\item [2] This tuple represents the input features, channels, and kernel size for the depthwise convolution.
\end{tablenotes}
\end{threeparttable}
\vspace{-1em}
\end{table}



\subsection{Comparison with Non-SIMD Convolutions}

As described in Section~\ref{sec:background}, non-SIMD-based techniques such as Cheetah~\cite{huang2022cheetah} achieve speedup by avoiding $\textbf{HRot}$ during HE convolution.
By eliminating $\textbf{HRot}$ operations, if we only take the computation latency of HE convolution into account, Cheetah is generally faster than Hyena for many standard convolution layers. 
However, this does not necessarily result in lower end-to-end latency since each HE convolution layer in hybrid PI protocols requires client-side encryption and decryption, cloud-side HE computation, and communicating input/output ciphertexts.
Since non-SIMD-based techniques do not support output channel packing in a ciphertext, the output ciphertext size after HE convolution becomes larger, thereby increasing the communication cost and putting more decryption burden on the client in return for reduced cloud-side computation.
To quantify this, we compare the end-to-end HE convolution latency, communication costs, and required storage capacity between Cheetah and Hyena in Table~\ref{tab:cheetah}.
Thanks to reduced client-side computation and communication costs for Hyena, the latency and communication cost are 1.5-2.5$\times$, 1.7$\times$, and storage is about 1.1-715$\times$ more efficient, respectively. 


\begin{figure}[t]
\centering
\includegraphics[width=0.88\columnwidth]{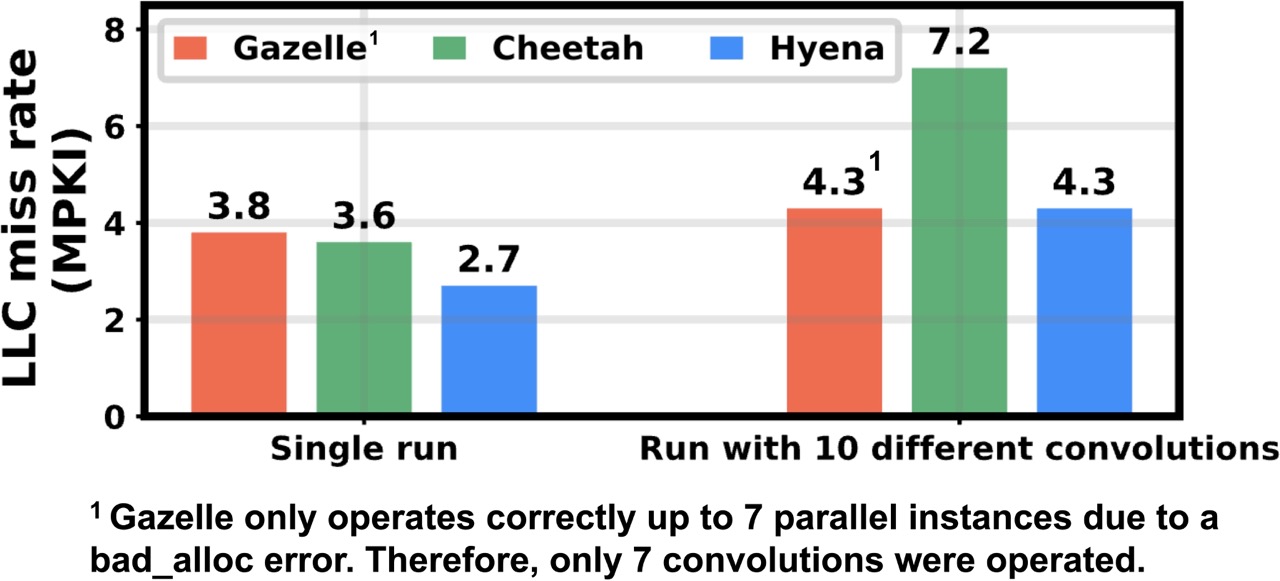}
\vspace{0em}
\caption{LLC miss rate (MPKI) when running convolution ($32, 2, 32,3$) on CPU. The tuple represents the input features ($H$=$W$), input channel count, output channel count, and kernel size ($f_w$=$f_h$).}
\vspace{-1em}
\label{fig:llc}
\end{figure}

\subsubsection{Sensitivity Study}
As shown in Fig.~\ref{fig:llc}, HE convolution has high last-level cache (LLC) miss rate.
Due to a large amount of required memory capacity (e.g., \ignore{model weights and switching keys}model weights, public keys, and input/output ciphertexts), LLC miss rate is 3.8\,MPKI in Gazelle.
Cheetah does not require switching keys for {\bf HRot}, but it also needs the model weights to be stored in plaintexts and has large output ciphertexts due to lack of output channel packing, showing slightly lower LLC miss rate than Gazelle.
In contrast, Hyena has the lowest LLC miss rate of 2.7\,MPKI.

In practical cloud environments, for high resource utilization and throughput, cloud does not process a single model inference; rather, multiple inferences are processed simultaneously (i.e., processing multiple HE convolutions in parallel) on a single machine, which leads to large memory access stalls.
When running the same HE convolution on top of 10 different large input-output channel convolutions, LLC miss rate in Cheetah and Hyena increases by 1.6-2$\times$ compared to single run. 
Due to limited resource, for Gazelle, the number of HE convolution running in parallel is limited to 7.
Therefore, in a practical cloud environment where multiple HE convolutions are executed in parallel, Hyena is the most sustainable due to the memory savings.

\ignore{\fixme{To assess how efficiently the convolution works in the high memory demands, 60 virtual memory workers were created to allocate and repetitively access a total of 360\,GB, thereby imposing significant stress on the memory system. 
In this high-load scenario, each convolution was repeated 200 times to measure the latency corresponding to the lower 25th percentile, which is depicted in Fig.~\ref{fig:memory}. 
Under these memory-constrained conditions, Gazelle showed the greatest spread with a standard deviation of 0.620 seconds, followed by Cheetah at 0.266 seconds and Hyena at 0.252 seconds.
}}

The end-to-end latency for a (32, 2, 32, 3) convolution, which includes client-side encryption and decryption, cloud-side HE computation, and communicating input/output ciphertexts, is compared in Fig.~\ref{fig:cdf}.
We assume the client utilizes 8\,GB of memory, and a network has a bandwidth of 100\,MB/s and a delay of 1\,ms. 
Each convolution with a single run for Hyena, Cheetah and Gazelle shows latency of 35.1, 35.6, and 48.9 seconds, respectively.
However, when running concurrently with other convolutions, the mean latency for Hyena and Cheetah deteriorates to 36.3, 38.7 with 10 convolutions (see Fig.~\ref{fig:cdf}(a)) and, for Gazelle, 57.8 seconds with 7 convolutions (see Fig.~\ref{fig:cdf}(b)), respectively.
The standard deviations are 0.27, 0.43, and 1.36 for Hyena, Cheetah, and Gazelle, respectively, indicating that Hyena shows the least latency variation.

\begin{figure}[t]
\centering
\includegraphics[width=0.8\columnwidth]{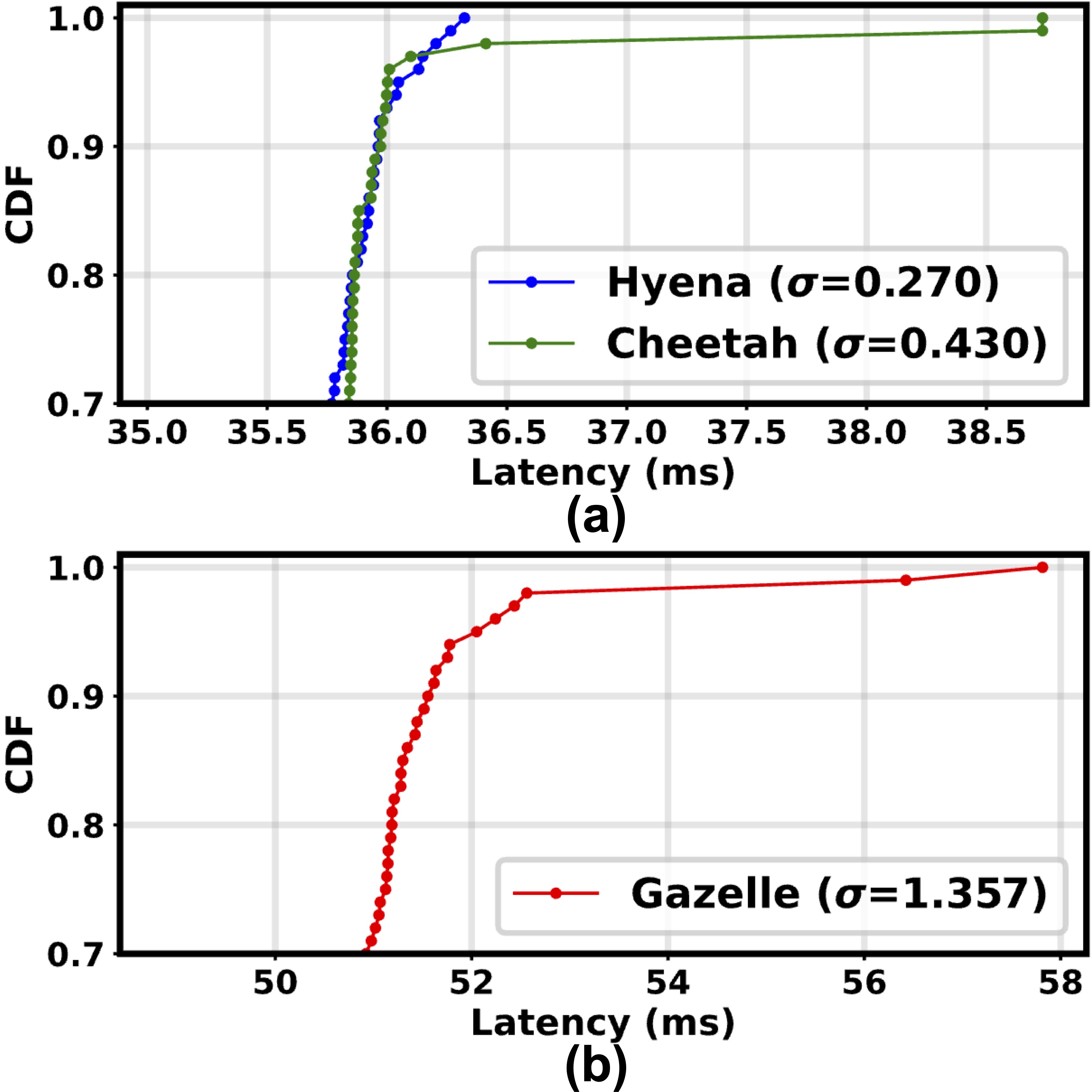}
\vspace{-1em}
\caption{Latency CDF (cumulative distribution function) for the worst 30\,\% of 100 end-to-end latency of ($32, 2, 32, 3$) on a network bandwidth of 100\,MB/s and a delay of 1\,ms with: (a) 10 and (b) 7 simultaneous convolutions.\ignore{: comparison of convolution~\cite{huang2022cheetah,juvekar2018gazelle} and proposed convolution.}}
\vspace{-1em}
\label{fig:cdf}
\end{figure}

\section{Conclusion}

This paper presents Hyena, a novel homomorphic convolution algorithm, improving the conventional algorithm in speed, communication cost, and weight storage efficiency. 
By employing a new plaintext multiplication algorithm based on the Walsh-Hadamard matrix, Hyena enables output channel packing, thereby reducing the communication cost and the weight storage.
Further optimization techniques, which involve optimal parameter selection and lazy reduction, lead to significant performance gains, achieving 1.6-3.8$\times$ faster processing and a 2000-8000$\times$ reduction in weight storage.
Compared to the conventional method, VGG-16, ResNet-20, and MobileNetV1 on ImageNet, employing Hyena, exhibit {1.3-2.5}$\times$ speedup, 2.1-7.9$\times$ memory saving, and {1.4-1.5}$\times$ communication cost saving.
\begin{acks}
    This work was supported in part by Institute of Information \& communications Technology Planning \& Evaluation (IITP) under the artificial intelligence semiconductor support program to nurture the best talents (IITP-2023-RS-2023-00256081) grant funded by the Korea government (MSIT), by National R\&D Program through the National Research Foundation of Korea (NRF) funded by Ministry of Science and ICT (RS-2024-00402495), and by Samsung Electronics Co., Ltd.
    Woo-Seok Choi is the corresponding author.
\end{acks}
\bibliographystyle{ACM-Reference-Format}
\bibliography{main}

\end{document}